\documentclass[english]{article}
\usepackage[T1]{fontenc}
\usepackage[latin9]{inputenc}
\usepackage{geometry}
\usepackage{color}
\usepackage{float}
\usepackage{amssymb}
\usepackage{graphicx}

\makeatletter
\date{}
\providecommand{\tabularnewline}{\\}

\usepackage{babel}
\usepackage{times}

\usepackage{epstopdf}
\usepackage{epsfig}

\usepackage{cite}

\makeatother

\usepackage{babel}
\begin{document}

\title{{Quantum Conference}}

\author{{Anindita Banerjee$^{a,}$}\thanks{email: anindita.phd@gmail.com}{,
Kishore Thapliyal$^{b,}$}\thanks{email: tkishore36@yahoo.com}{,
Chitra Shukla$^{c,}$}\thanks{email: shukla.chitra@i.mbox.nagoya-u.ac.jp }{,
Anirban Pathak$^{b,}$}\thanks{email: anirban.pathak@jiit.ac.in}\\
$^{a}$Department of Physics and Center for Astroparticle
Physics and Space Science,\\ Bose Institute, Block EN, Sector V, Kolkata
700091, India \\
$^{b}$Jaypee Institute of Information Technology, A-10,
Sector-62, Noida, UP-201307, India \\
$^{c}$Graduate School of Information Science, Nagoya
University, Furo-cho 1, Chikusa-ku,\\ Nagoya, 464-8601, Japan}
\maketitle
\begin{abstract}
{\normalsize{}A notion of quantum conference is introduced in analogy
with the usual notion of a conference that happens frequently in today's
world. Quantum conference is defined as a multiparty secure communication
task that allows each party to communicate their messages simultaneously
to all other parties in a secure manner using quantum resources. Two
efficient and secure protocols for quantum conference have been proposed.
The security and efficiency of the proposed protocols have been analyzed
critically. It is shown that the proposed protocols can be realized
using a large number of entangled states and group of operators. Further,
it is shown that the proposed schemes can be easily reduced to protocol
for multiparty quantum key distribution and some earlier proposed
schemes of quantum conference, where the notion of quantum conference
was different.}{\normalsize \par}
\end{abstract}

\section{{\normalsize{}Introduction\label{sec:Introduction}}}

In 1984, an unconditionally secure key distribution protocol using
quantum resources was proposed by Bennett and Brassard \cite{bb84}.
The scheme, which is now known as BB84 protocol drew considerable
attention of the cryptography community by its own merit as it offered
unconditional security, which was unachievable by any classical protocol
of key distribution. However, the relevance of BB84 quantum key distribution
(QKD) protocol and a set of other schemes of QKD were actually established
very strongly in 1994, when the seminal work of Shor \cite{Shor-algo}
established that RSA \cite{RSA} and a few other schemes of classical
cryptography \cite{DF} would not remain secure if a scalable quantum
computer is built. The BB84 protocol, not only established the possibility
of obtaining unconditional security, but also manifested enormous
power of quantum resources that had been maneuvered since then. Specifically,
this attempt at the unconditional security of QKD was followed by
a set of protocols for the same task \cite{ekert,b92,vaidman-goldenberg}.
Interestingly, the beautiful applications of quantum mechanics in
secure communication did not remain restricted to key distribution.
In fact, it was realized soon that the messages can be sent in a secure
manner without preparing a prior key \cite{ping-pong}. Exploiting
this idea various such schemes were proposed which fall under the
category of secure direct quantum communication (\cite{ping-pong,dsqc-1,review,Anindita,cdsqc}
and references therein).

The schemes for secure direct quantum communication can be categorized
into two classes on the basis of additional classical communication
required by the receiver (Bob) to decode each bit of the transmitted
message- (i) quantum secure direct communication (QSDC) \cite{ping-pong,dsqc-1,review}
and (ii) deterministic secure quantum communication (DSQC) \cite{Anindita}.
In the former, Bob does not require an additional classical communication
to decode the message, while such a classical communication is involved
in the latter (see \cite{book} for review). It is worth noting that
in a scheme of QSDC/DSQC meaningful information flows in one direction
as it only allows Alice to send a message to Bob in an unconditionally
secure manner using quantum resources and without generation of a
key. However, in our daily life, we often require two way communication
(say, when we speak on a telephone). Interestingly, a modification
of one of the first few QSDC schemes (i.e., ping-pong scheme \cite{ping-pong})
led to a new type of protocol that allows both Alice and Bob to communicate
simultaneously using the same quantum channel. This scheme for simultaneous
two way communication was first proposed by Ba An \cite{ba-an} and
is known as quantum dialogue (QD). Due to its similarity with the
task performed by telephones, a scheme for QD are also referred as
quantum telephone \cite{quantum telephon1,telephone} or quantum conversation\footnote{It may be noted that in an ideal scheme of QD, information encoded
by two parties exist simultaneously in a channel, but in the protocol
for quantum conversation introduced in \cite{sakshi-panigrahi-epl},
it was not the case. However, the communication task in hand was equivalent.} \cite{sakshi-panigrahi-epl} scheme, but in what follows, we will
refer to them as QD. Due to its practical relevance, schemes of QD
received much attention and several new schemes of QD have been proposed
in the last decade \cite{QD-EnSwap,QD-GHZ,QD-hwang,QD-qutrit}. However,
all these schemes of QD, and also the schemes of QSDC and DSQC, mentioned
here are restricted to the two-party scenario. This observation led
to two simple questions- (i) Do we need a multiparty QD for any practical
purpose? and (ii) If answer of the previous question is yes, can we
construct such a scheme? It is easy for us (specially for the readers
of this paper and the authors of the similar papers who often participate
in conferences and meet as members of various committees) to recognize
that conferences and meetings provide examples of situation where
multiparty dialogue happens. Specifically, in a conference a large
number of participants can exchange their thoughts (inputs, which
may be viewed as classical information). Although, usually participants
of the conference/meeting are located in one place, but with the advent
of new technologies, tele-conferences, webinar, and similar ideas
that allow remotely located users to get involved in multiparty dialogue,
are becoming extremely popular. For the participants of such a conference
or meeting that allows users to be located at different places, desirable
characteristics of the scheme for the conference should be as follows-
(A) A participant must be able to communicate directly with all other
participants, or in other words, every participant must be able to
listen the talk/opinion delivered by every speaker as it happens in
a real conference. (B) A participant should not be able to communicate
different opinion/message to different users or user groups. (C) Illegitimate
users or unauthorized parties (say those who have not paid conference
registration fees) will not be able to follow the proceedings of the
conference. It is obvious that criterion (C) requires security and
a secure scheme for multiparty quantum dialogue satisfying (A)-(C)
is essential for today's society. We refer to such a scheme for multiparty
secure communication that satisfies (A)-(C) as a\textcolor{red}{{} }scheme
for quantum conference (QC) because of its analogy with the traditional
conferences (specially with the tele-conferences). The analogy between
the communication task performed here and the traditional conference
can be made clearer by noting that Wikipedia defines conference as
``a conference is a meeting of people who \textquotedbl{}confer\textquotedbl{}
about a topic'' \cite{Wiki}. Similarly, Oxford dictionary describes
a conference as ``a linking of several telephones or computers, so
that each user may communicate with the others simultaneously'' \cite{Ox-Dic}.
This is exactly the task that the proposed protocol for QC is aimed
to perform using quantum resources and in a secure manner. Thus, QC
is simply a conference, which is an $n$-party communication, where
each participant can communicate his/her inputs (classical information)
using quantum resources to remaining $(n-1)$ participants. However,
it should be made clear that it is neither a multi-channel QSDC nor
a multi-channel QD scheme. To be precise, one may assume that each
participant maintains private quantum channels with all other participants
and uses those to communicate his/her input to others via QSDC or
QD. This is against the idea of a conference, as in this arrangement,
a participant may send different information/opinion to different
participants, in violation of Criterion (B) listed above. The fact
that to the best of our knowledge, no such scheme for multiparty secure
quantum communication exists has motivated us to introduce the notion
of QC and to aim to design a scheme for the same.

Here it would be apt to note that although no scheme for QC is yet
proposed, various schemes for other multiparty quantum communication
tasks have already been proposed. For example, quantum schemes for
voting \cite{TZLcom}, auction \cite{auction,Our-auction}, and e-commerce
\cite{online-shop} are necessarily expected to be multiparty quantum
communication schemes. Interestingly, there are a few schemes for
all these tasks proposed in the past (\cite{TZLcom,online-shop,auction,Our-auction}
and references therein). Another recently discussed multiparty task
is\textcolor{red}{{} }quantum key agreement (QKA) (\cite{QKA-cs} and
references therein), where the final key is generated by the contribution
of all the parties involved, and a single or a few parties can not
decide the final key. For instance, a multiparty QKA scheme \cite{QKA-cs}
was proposed in the recent past, in which encoded qubits travel in
a circular manner among all the parties. In fact, most of these multiparty
quantum communication schemes, except QKA, can be intrinsically viewed
as a (many) sender(s) sending some useful information in a secure
manner to a (many) receiver(s) under the control of a third party.
Further, all these schemes can be broadly categorized as secure multiparty
quantum communication and secure multiparty quantum computation. Though
the line between the two is very faint to distinguish and categorize
a scheme among one of them, QKA and e-commerce may be considered in
the former, while voting and auction fall under the latter. Some efforts
have also been made to introduce a notion of QC as a multiparty quantum
communication task. However, earlier ideas of QC can be viewed as
special cases of the notion of QC presented here and they are not
sufficient to perform a conference as defined above in analogy with
the definition provided in Oxford dictionary and other sources.

Bose, Vederal and Knight \cite{bose} proposed a generalized entanglement-swapping-based
scheme for multiparty quantum communication that led to a set of quantum
communication schemes related to QC, viz., cryptographic conference
\cite{bose}, conference key agreement and conference call \cite{chitra},
and a scheme where many senders send their messages to single receiver
via generalized superdense coding \cite{bose}. In cryptographic conference,
all parties share a multipartite entangled state. They perform measurement
in the computational or diagonal basis, and the results of those measurements
in which the bases chosen by all the users coincide are used to establish
the secret key which will be known to all the users within the group.
A similar notion of conference key agreement was used in \cite{chitra},
where a generalized notion of dense coding was used. Clearly the notion
of conference is weaker here, and in our version of conference such
keys can be distributed easily if all the users communicate random
bits instead of meaningful messages. Recent success of designing the
above mentioned schemes for multiparty quantum communication further
motivated us to look for a scheme for QC.

A two party analogue of QC can be considered as QD, where both parties
can communicate simultaneously. The group theoretic structure of Ba-An-type
QD schemes has been discussed in Ref. \cite{qd}. The group theoretic
structure discussed in \cite{qd,QKA-cs,AQD} will be exploited here
to introduce the concept of QC. Further, an asymmetric counterpart
of the Ba-An-type QD scheme is proposed in the recent past \cite{AQD}.
Following which we will also introduce and briefly discuss an asymmetric
QC (AQC), where all the parties involved need not to send an equal
amount of information. With the recent interest of quantum communication
community on quantum internet \cite{QInternet,QInternet2} and experimental
realization of multiparty quantum communication schemes \cite{Exp-multiparty},
the motivation for introducing a QC or AQC scheme can be established.

Remaining part of the paper is organized as follows. Sec. \ref{sec:Ba-An-protocol}
is dedicated to a brief review of QD and the group theoretic approach
of QD for the sake of completeness of the paper, which has been used
in the forthcoming sections to develop the idea of QC. Two general
schemes for the task of QC have been introduced in Sec. \ref{sec:Quantum-conferencePro}.
In the next section, we have considered a few specific examples of
both these schemes. The feasibility of an AQC scheme has also been
discussed in Sec. \ref{sec:Examples-and-possible}. Finally, the security
and efficiency of the proposed schemes have been discussed in Sec.
\ref{sec:Security-analysis} before concluding the paper in Sec. \ref{sec:Conclusion}.

\section{\textcolor{black}{\normalsize{}Ba An protocol of QD and its generalization
using modified Pauli group \label{sec:Ba-An-protocol}}}

It would be relevant to mention that some of the present authors had
presented the general structure of QD protocols in \cite{qd} and
established that the set of unitary operators used by Alice and Bob
must form a group under multiplication. The group structure has also
been found to be suitable for the asymmetric QD schemes \cite{AQD},
where Alice and Bob use encoding operations from different subgroups
of a modified Pauli group, like $G_{1}=\{I,\,X,\,iY,\,Z\}$. This
particular Abelian group ($G_{1}$) is of order 4 under multiplication
and is called a modified Pauli group as we neglect the global phase
in the product of any two elements of this group, which is consistent
with the quantum mechanics (for detail see \cite{qd,AQD}). The generalized
group $G_{n}$ can be formed by $n$-fold tensor products of $G_{1}$,
i.e., $G_{n}=G_{1}^{\otimes n}$. In the original QD protocol \cite{ba-an},
the encoding is done by Alice and Bob, respectively, using the same
set of operations $\left\{ U_{i}\right\} $ from the modified Pauli
group $G_{1}$. The entire scheme of Ba An \cite{ba-an} can be summed
up in the formula $|\psi_{j}\rangle_{final}=$ $U_{B}U_{A}|\psi_{i}\rangle_{initial}:_{initual}\langle\psi_{i}|\psi_{j}\rangle_{final}=\delta_{i,j}$,
where $|\psi_{k}\rangle$ are the Bell states. It is required that
all the possible final states obtained after Alice's and Bob's encoding
operations should remain orthonormal to each other and also with the
initial state. Once the initial and final states are known to both
the legitimate users, they can exploit knowledge of their own encoding
operation to extract each other's message.

Interestingly, Alice and Bob encode information with the same operators,
say, $I$ for 00, $X$ for 01, $iY$ for 10, and $Z$ for 11. In this
scenario,\textcolor{blue}{{} }Alice obtains a unique bijective mapping
from the composite encoding of Alice and Bob ($U_{B}.U_{A}$) to Bob's
operation ($U_{B}$) using her unitary operation ($U_{A}$). This
is obvious where there are only 2 parties, we may ask, is it possible
to extend this scheme for QD to design a scheme for multiparty conference?
Let us examine two cases with 3 parties: in Case 1: when all the parties
encode the same bits say, 00 i.e., they apply $U_{C_{I}},\,U_{B_{I}}$
and $U_{A_{I}}$; and in Case 2: when one of them encodes the same
bits used in Case 1, i.e., 00 and other two will encode the similar
bits but other than 00, say 01 $(\mbox{or }10,\,11)$, i.e., they
apply $U_{C_{X}},\,U_{B_{X}}$ and $U_{A_{I}}$, respectively. In
these two cases, the resultant state is always the same as what was
prepared initially, and none of the parties can deterministically
conclude each others encoding. In fact, there will be many such cases,
hence, Ba An's original protocol for QD cannot be generalized directly
to design a scheme for multiparty conference. 

To design a scheme for QC, we will use the idea of disjoint subgroups
introduced by some of the present authors in the recent past \cite{QKA-cs}.
Disjoint subgroups refer to subgroups, say $g_{i}$ and $g_{j}$,
of a group $G_{n}$ such that they satisfy $g_{i}\cap g_{j}=\{I\}$.
Thus, except Identity $g_{i}$ and $g_{j}$ do not contain any common
element. The modified Pauli group $G_{1}$ has 3 mutually disjoint
subgroups: $g_{1}=\{I,\,X\},$ $g_{2}=\{I,iY\}$ and $g_{3}=\{I,\,Z\}$.
Whenever there are more than two parties, we can encode using disjoint
subgroups of operators, i.e., each party may be allowed to encode
with a unique disjoint subgroup. For example, if Alice, Bob and Charlie
want to set up a QC among them, then Alice can encode using $g_{1}$,
Bob can encode using $g_{2}$ and Charlie can encode using $g_{3}.$
The use of disjoint subgroups circumvents the limitations of the original
two-party QD scheme and provides a unique mapping required for multiparty
conversation.

In what follows, we have proposed two protocols to accomplish the
task of a QC scheme.

\section{\textcolor{black}{\normalsize{}Quantum conference \label{sec:Quantum-conferencePro}}}

Here, we have designed two multiparty quantum communication schemes
where prior generation of key is not required. These schemes may be
used for QC, i.e., for multiparty communication of meaningful information
among the users. Additionally, it is easy to observe that these schemes
naturally reduce to the schemes for multiparty key distribution if
the parties send random bits instead of meaningful messages.

\subsection{{\normalsize{}Protocol 1: Multiparty QSDC scheme for QC}}

Let us start with the simplest case, where $\left(N-1\right)$ parties
send their message to $N$th party. This can be thought of as a multiparty
QSDC. Suppose all the parties decide to encode or communicate $k$-bit
classical messages. In this case, each user would require a subgroup
of operators with at least $2^{k}$ operators. In other words, each
party would need at least a subgroup $g_{i}$ of order $2^{k}$ of
a group $\mathcal{G}$. Here, we would like to propose one such multiparty
QSDC scheme. 
\begin{description}
\item [{Step~1.1}] First party Alice be given one subgroup $g_{A}=\left\{ A_{1},A_{2},\ldots,A_{2^{k}}\right\} $
to encode her $k$-bit information. Similarly, other parties (say
Bob and Charlie) can encode using subgroups $g_{B}=\left\{ B_{1},B_{2},\ldots,B_{2^{k}}\right\} $,
and $g_{C}=\left\{ C_{1},C_{2},\ldots,C_{2^{k}}\right\} $, and so
on for $\left(N-1\right)$th party Diana,\textcolor{red}{{} }whose encoding
operations are $g_{D}=\left\{ D_{1},D_{2},\ldots,D_{2^{k}}\right\} $.
\\
 All these subgroups are pairwise disjoint subgroups, i.e., they are
chosen in such a way that $g_{i}\cap g_{j}=\left\{ \mathbb{I}\right\} \forall i,j\in\{1,2,\cdots,N-1\}$.
As the requirement for encoding operations to be from disjoint subgroups
has been already established beforehand.\\
 Additionally, here we assume that all the parties do nothing (equivalent
to operator Identity) on their qubits for encoding a string of $k$
zeros. As Identity is the common element in the set of encoding operations
to be used by each party it will be convenient to consider this as
a convention in the rest of the paper. 
\item [{Step~1.2}] Nathan (the $N$th party) prepares an $n$-qubit entangled
state $|\psi\rangle$ (with $n\ge\left(N-1\right)k$). \\
 It is noteworthy that maximum information that can be encoded on
the $\left(N-1\right)k$-qubit quantum channel is $\left(N-1\right)k$
bits and here $\left(N-1\right)$ parties are sending $k$ bits each.
In other words, after encoding operation of all the $\left(N-1\right)$
parties the quantum states should be one of the $2^{\left(N-1\right)k}$
possible orthogonal states. 
\item [{Step~1.3}] Nathan sends $m$ qubits ($m<n$) of the entangled
state $|\psi\rangle$ to Alice in a secure manner, who applies one
of the operations $A_{i}$ (which is an element of the subgroup of
operators available with her) on the travel qubits to encode her message.
This will transform the initial state to $|\psi_{A}\rangle=A_{i}|\psi\rangle$.
Subsequently, Alice sends all these encoded qubits to the next user
Bob. 
\item [{Step~1.4}] Bob encodes his message which will transform the quantum
state to $|\psi_{B}\rangle=B_{j}A_{i}|\psi\rangle$. Finally, he also
sends the encoded qubits to Charlie in a secure manner. 
\item [{Step~1.5}] Charlie would follow the same strategy as followed
by Alice and Bob. In the end, Diana receives all the encoded travel
qubits and she also performs the\textcolor{red}{{} }operation corresponding
to her message to transform the state into $|\psi_{i,j,k,\ldots,l}^{\prime}\rangle=D_{l}\cdots C_{k}B_{j}A_{i}|\psi\rangle$.
She returns all the travel qubits to Nathan. 
\item [{Step~1.6}] Nathan can extract the information sent by all $\left(N-1\right)$
parties by measuring the final state using an appropriate basis set.\\
 It may be noted that Nathan can decode messages sent by all $\left(N-1\right)$
parties, if and only if the set of all the encoding operations gives
orthogonal states after their application on the quantum state, i.e.,
$\left\{ |\psi_{i,j,k,\ldots,l}^{\prime}\rangle\right\} $ are orthogonal
for all $i,j,k,\ldots,l\in\left\{ 1,\cdots2^{k}\right\} $. In other
words, after the encoding operation of all the $\left(N-1\right)$
parties the quantum states should be a part of a basis set with $2^{\left(N-1\right)k}$
orthogonal states for unique decoding of all possible encoding operations. 
\end{description}
This scheme can be viewed as the generalization of ping-pong protocol
\cite{ping-pong} to a multiparty scenario, where multiple sender's
can simultaneously send their information to a receiver. In a similar
way, if all the senders wish to send and receive the same amount of
information, then all of them can also choose to prepare their initial
state $|\psi\rangle$ independently and send it to all other parties
in a sequential manner. Subsequently, all of them may follow the above
protocol faithfully to perform $N$ simultaneous multiparty QSDC protocols.

In fact, $N$ simultaneous multiparty QSDC schemes of the above form
will perform the task required in an ideal QC scheme. However, as
each sender has to encode his secret multiple times ($N-1$ times),
it would allow him to encode different information in each round.
Though it may be advantageous in some communication schemes, where
a sender is allowed to send different bit values to different receivers,
but is undesirable in a scheme for QC. Specifically, to stress on
the relevance of a scheme that allows each sender to encode different
bits to all the receivers, we may consider a situation where each
party (or a few of them) publicly asks a question, and the receivers
answer the question independently (for an analogy think of a panel
discussion in television). In this case, all the receivers may have
different opinions (say one may agree with some of them and may not
with the remaining) about various questions being asked. As far as
a scheme for QC is concerned, Protocol 1 described here would work
under the assumption of semi-honesty. Specifically, a semi-honest
party may try to cheat, but he/she would follow the protocol faithfully.
This assumption would enable us to consider that each party is encoding
the same information every time. In what follows. we will establish
that such an assumption is not required. Specifically, in Protocol
2, we aim to design a genuine QC scheme, which does not require the
semi-honesty assumption to restrict a user from sending different
information to different receivers.

\subsection{{\normalsize{}Protocol 2: Multiparty QD-type scheme for QC}}

Here, we will attempt to design an efficient QC scheme, which can
be thought of as a generalized QD scheme. In analogy of the original
Ba-An-type QD scheme, we will need the set of encoding operations
for the $N$th party (Nathan). Here, firstly we propose the protocol
which is followed by a prescription to obtain the set of operations
for $N$th party, assuming a working scheme designed for the Protocol
1. 
\begin{description}
\item [{Step~2.1}] Same as that of Step 1.1 of Protocol 1 with a simple
modification that also provide Nathan a subgroup $g_{N}=\left\{ N_{1},N_{2},\ldots,N_{2^{k}}\right\} $which
enables him to encode a $k$-bit message at a later stage. \\
 The mathematical structure of this subgroup will be discussed after
the protocol. 
\item [{Step~2.2}] Same as Step 1.2 of Protocol 1. 
\item [{Step~2.3}] Same as Step 1.3 of Protocol 1. 
\item [{Step~2.4}] Same as Step 1.4 of Protocol 1. 
\item [{Step~2.5}] Same as Step 1.5 of Protocol 1. 
\item [{Step~2.6}] Nathan applies unitary operation $N_{m}$ to encode
his secret and the resulting state would be $|\psi_{i,j,k\cdots l,m}^{\prime\prime}\rangle=N_{m}D_{l}\cdots$ $ C_{k}B_{j}A_{i}|\psi\rangle$. 
\item [{Step~2.7}] Nathan measures $|\psi_{i,j,k\cdots l,m}^{\prime\prime}\rangle$
using the appropriate basis as was done in Step 1.6 of Protocol 1
and announces the measurement outcome. Now, with the information of
the initial state, final state and one's own encoding all parties
can extract the information of all other parties.\\
 It is to be noted that the information can be extracted only if the
set of all the encoding operations gives orthogonal states after their
application on the quantum state, i.e., all the elements of $\left\{ |\psi_{i,j,k\cdots l,m}^{\prime\prime}\rangle\right\} $
are required to be mutually orthogonal for $i,j,k\cdots l,m\in\left\{ 1,\cdots2^{k}\right\} $.
In other words, after the encoding operation of all the $N$ parties
the set of all possible quantum states should form a $2^{\left(N-1\right)k}$
dimensional basis set. 
\end{description}
Nathan's unitary operation can be obtained using the fact that the
remaining $\left(N-1\right)$ parties have already utilized the channel
capacity. Hence, his encoding should be in such a way that after his
encoding operation $N_{m}$, the final quantum state should remain
an element of the basis set in which the initial state was prepared.
However, the bijective mapping between the initial and final states
present in Protocol 1 would disappear here. This is not a limitation.
It is actually a requirement. This is so because, in contrast to Protocol
1 where the initial and final states are secret, in Protocol 2, the
choice of the initial state and the final state are publicly broadcasted.
Existence of a bijective mapping would have revealed all the secrets
to Eve.  This condition provides us a mathematical advantage. Specifically,
it allows us to construct the set of unitary operations that Nathan
can apply. To do so we need to use the information about the disjoint
subgroups of operators that are used by other parties. The procedure
for construction of Nathan's set of operations is described below.

For simplicity, let us write the encoding operations of all the parties
as follows:

\[
\begin{array}{ccccc}
 & \tilde{0} & \tilde{1} & \cdots & \tilde{\left(2^{k}-1\right)}\\
{\rm Alice} & A_{1} & A_{2} & \cdots & A_{2^{k}}\\
{\rm Bob} & B_{1} & B_{2} & \cdots & B_{2^{k}}\\
{\rm Charlie} & C_{1} & C_{2} & \cdots & C_{2^{k}}\\
\vdots & \vdots & \vdots & \vdots & \vdots\\
{\rm Diana} & D_{1} & D_{2} & \cdots & D_{2^{k}}\\
{\rm Nathan} & N_{1} & N_{2} & \cdots\cdots & N_{2^{k}}.
\end{array}
\]
Here, $\tilde{x}$ corresponds to the binary value of the decimal
number $x$, and it represents the classical information to be encoded
by user X{*}{*}{*} (listed in Column 1) using the the operator $X_{x+1}$ (listed
in $x+1$th column in the row corresponding to the user X{*}{*}{*}).
For example, to encode $01=\tilde{1},$ Alice would use the operator
$A_{1+1}=A_{2}$, whereas for the same encoding Bob and Charlie would
use $B_{2}$ and $C_{2}$, respectively. Further, we would like to
note that by construction operators $X_{i}=X_{i}^{-1}$ as $X_{i}$
is an element of the modified Pauli group, and it is assumed that
the encoding operations of the different users are chosen from the disjoint
subgroups of the modified Pauli groups in such a way that the product
of operations listed in any column is Identity, i.e., 
\begin{equation}
A_{i}B_{i}C_{i}\cdots D_{i}N_{i}=\mathbb{I}\,\forall i.\label{eq:condition}
\end{equation}
This implies that if all the parties encode the same secret then the
final state and the initial state would be the same. To illustrate
this we may consider following example 
\begin{equation}
\begin{array}{ccc}
 & \tilde{0} & \tilde{1}\\
{\rm Alice} & \mathbb{I} & X\\
{\rm Bob} & \mathbb{I} & iY\\
{\rm Nathan} & \mathbb{I} & Z.
\end{array}\label{eq:example}
\end{equation}

From Eqs. (\ref{eq:condition})-(\ref{eq:example}), it is clear that
the choice of encoding operations of the other users (i.e., $A_{i}.B_{i},C_{i},\cdots,D_{i})$
would uniquely determine $N_{i}$. Further, it is assumed that the
encoding operations used by different users to encode $\tilde{x}$
are selected in a particular order that ensures $X_{i}X_{j}=X_{k}\forall X$
and particular choice of $i,j.$ For example, this condition implies
that if Alice's operators satisfy $A_{2}A_{3}=A_{5},$ then Bob and
Charlie would be given the encoding operators in an order that satisfy
$B_{2}B_{3}=B_{5}$ and $C_{2}C_{3}=C_{5},$ respectively, and the
same ordering of operators will be applicable to all other users.
Now, using the above mentioned facts and convention, we need to establish
that $\left\{ N_{x}\right\} $ forms a group under multiplication.
\textcolor{red}{Eq.} (\ref{eq:condition})\textcolor{red}{{} }and the self reversibility
of the elements $X_{i}$ lead to following identity- $N_{i}=D_{i}^{-1}\cdots C_{i}^{-1}B_{i}^{-1}A_{i}^{-1}=D_{i}\cdots C_{i}B_{i}A_{i}.$
This may be used to establish the closure property of the group $\left\{ N_{x}\right\} $
as $N_{i}N_{j}=\left(D_{i}\cdots C_{i}B_{i}A_{i}\right)\left(D_{j}\cdots C_{j}B_{j}A_{j}\right)=\left(D_{i}D_{j}\right)\cdots\left(C{}_{i}C_{j}\right)\left(B_{i}B_{j}\right)\left(A_{i}A_{j}\right)=\left(D_{k}\cdots C_{k}B_{k}A_{k}\right)=N_{k}\in\left\{ N_{x}\right\} $.
This is so because the Pauli operators commute with each other under
the operational definition of multiplication used in defining the
modified Pauli group. All the remaining properties of the group follows
directly from the nature of Pauli operators used to design $X_{x}$.
Thus, it is established that the generalized multiparty QSDC scheme
can be modified to a generalized QD scheme. It will be interesting
to obtain the original Ba An's QD scheme as a limiting case as follows.
\[
\begin{array}{ccccc}
 & \tilde{0} & \tilde{1} & \tilde{2} & \tilde{3}\\
{\rm Alice} & \mathbb{I} & X & iY & Z\\
{\rm Bob} & B_{1} & B_{2} & B_{3} & B_{4}.
\end{array}
\]
This particular case and all the discussions leave us with $\left\{ B_{i}\right\} =\left\{ \mathbb{I},X,iY,Z\right\} ,$
which is identical with Alice's operations.

In Table \ref{tab:conference}, we have provided a list comprising
of the number of participants in the QC and the number of cbits they
want to encode. The table explicitly mentions different multipartite
states or quantum channels that can be utilized for the same.

\begin{center}
\begin{table}
\begin{centering}
\begin{tabular}{|c|c|c|c|c|}
\hline 
Number of  & cbits by  & Groups  & Number of  & Entangled states\tabularnewline
parties $\left(N\right)$  & each party$\left(k\right)$  &  & travel qubit $\left(m\right)$  & \tabularnewline
\hline 
3  & 1  & $G_{1}$  & 1  & Bell\tabularnewline
\hline 
4 & 1  & $G_{2}$  & 2  & 4-qubit cluster or $|\Omega\rangle$ state \tabularnewline
\hline 
4 & 1  & $G_{2}^{1}(8),G_{2}^{2}(8),G_{2}^{4}(8),G_{2}^{5}(8)$  & 2  & GHZ \tabularnewline
\hline 
5  & 1  & $G_{2}$  & 2  & 4-qubit cluster or $|\Omega\rangle$ state \tabularnewline
\hline 
2  & 2  & $G_{1}$ & 1  & Bell \tabularnewline
\hline 
3  & 2  & $G_{2}$  & 2  & 4-qubit cluster or $|\Omega\rangle$ state \tabularnewline
\hline 
2  & 3  & $G_{2}$  & 2  & 4-qubit cluster or $|\Omega\rangle$ state \tabularnewline
\hline 
2  & 3  & $G_{2}^{1}(8),G_{2}^{2}(8),G_{2}^{4}(8),G_{2}^{5}(8)$  & 2  & GHZ \tabularnewline
\hline 
2  & 4  & $G_{2}$ & 2  & 4-qubit cluster or $|\Omega\rangle$ state \tabularnewline
\hline 
\end{tabular}
\par\end{centering}

\protect\caption{\label{tab:conference} Various possibilities of QC scheme with a
maximum number of $N$ parties each encoding $k$ bits using a group
of unitary operators with at least $2^{\left(N-1\right)k}$ elements.
The quantum states suitable in each case and corresponding number
of travel qubits are also mentioned.}
\end{table}

\par\end{center}

Finally, it is also worth mentioning here that this protocol is free
from the individual participant's attack as each user is allowed to
encode only once. The remaining attack strategies and security against
them will be discussed in detail in Sec. \ref{sec:Security-analysis}.
Before doing so, it may be noted that the message is extracted in
different ways in Protocol 1 and 2. Specifically, in Protocol 1, the
encoding of each sender is inferred from the bijective mapping between
the initial and final states, in analogy to the QSDC protocols. In
Protocol 2, the same task is achieved by each party by exploiting
the bijective mapping between the final state and his/her own encoding,
which is analogous to the QD protocols. Therefore, Protocol 1 (2)
proposed here can be viewed as a generalized multiparty QSDC (QD)
scheme.

\section{{\normalsize{}Examples and possible modifications }\textcolor{black}{\normalsize{}\label{sec:Examples-and-possible}}}

\begin{figure}[H]
\centering{}\includegraphics{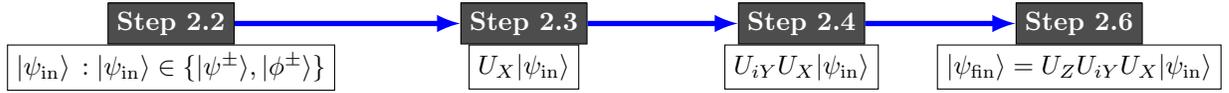}\protect\caption{\label{fig:example} The evolution of the quantum channel with all
the intermediate states and corresponding encoding in an example scheme
are summarized. Here, the unitary operations $U_{X}:\,U_{X}\in\left\{ \mathbb{I},X\right\} ,$
$U_{iY}\in\left\{ \mathbb{I},iY\right\} ,$ and $U_{Z}\in\left\{ \mathbb{I},Z\right\} .$ }
\end{figure}

Let's elaborate the proposed idea by discussing a particular example
of the proposed Protocol 2 for a 3 party case, where each party encodes
only one bit. To begin with, let us assume that one of the parties,
say, Nathan, prepares an entangled state in the Bell basis $|\psi_{{\rm in}}\rangle$
in Step 2.2 (the same is also illustrated through Figure \ref{fig:example},
where the quantum state transforming in the various intermediate steps
is mentioned). Nathan sends one of the qubits of the Bell state to
Alice in Step 2.3, who encodes a unitary operation $U_{X}:\,U_{X}\in\left\{ \mathbb{I},X\right\} $
corresponding to her secret and sends it to Bob. Similarly, Bob also
encodes his message in Step 2.4 using a unitary operation $U_{iY}:\,U_{iY}\in\left\{ \mathbb{I},iY\right\} $.
Finally, Nathan receives the encoded qubit in Step 2.5 and encodes
his message using a unitary operation $U_{Z}:\,U_{Z}\in\left\{ \mathbb{I},Z\right\} $
in Step 2.6. Finally, in Step 2.7, Nathan measures the final quantum
state $|\psi_{{\rm fin}}\rangle=U_{Z}U_{iY}U_{X}|\psi_{{\rm in}}\rangle$
in the Bell basis and announces the measurement outcome. From the
knowledge of the encoding operation performed by himself/herself,
and the initial and final Bell states all three parties would be able
to decode the secrets sent by the remaining two parties, for which
they have to use the bijective mapping present between his/her own
encoding and the pair of encoding operations performed by the other
two users. For instance, we may consider a particular case, where
Nathan's choice of the initial state and measurement outcomes are
the same, say $|\psi^{+}\rangle$. This announcement leaves only two
possibilities, either $U_{Z}=U_{iY}=U_{X}=\mathbb{I}$ or $U_{X}=X$,
$U_{iY}=iY$ and $U_{Z}=Z$. In this particular case, each party knows
whether they have encoded 0 (i.e., applied Identity) or not. Using
which they can extract the message sent by the remaining users. We
may further note that if we restrict Nathan to always encode Identity,
then this scheme (Protocol 2) will reduce to Protocol 1.

Further examples with the higher number of parties involved in the
QC are summarized in Table \ref{tab:Multiparty-Quantum-conference.}.
The examples listed are not the unique choices and similar set of
unitary operators may be easily obtained using the prescription defined
in the previous section.

\begin{table}
\begin{centering}
\begin{tabular}{|c|c|c|c|}
\hline 
Party  & c-bits  & Quantum state  & Set of encoding operations\tabularnewline
\hline 
3  & 1  & Bell or GHZ  & $\left\{ P_{1}:\left\{ \mathbb{I},X\right\} ,\,P_{2}:\left\{ \mathbb{I},iY\right\} ,\,P_{3}:\left\{ \mathbb{I},Z\right\} \right\} $\tabularnewline
\hline 
3  & 2  & 4-qubit cluster state  & $\left\{ P_{1}:\left\{ \mathbb{I}\otimes\mathbb{I},\mathbb{I}\otimes X,X\otimes\mathbb{I},X\otimes X\right\} ,\right.$\tabularnewline
 &  &  & $\left.P_{2}:\left\{ \mathbb{I}\otimes\mathbb{I},\mathbb{I}\otimes iY,iY\otimes\mathbb{I},iY\otimes iY\right\} ,\,P_{3}:\left\{ \mathbb{I}\otimes\mathbb{I},\mathbb{I}\otimes Z,Z\otimes\mathbb{I},Z\otimes Z\right\} \right\} $\tabularnewline
\hline 
4  & 1  & GHZ  & $\left\{ P_{1}:\left\{ \mathbb{I}\otimes\mathbb{I},X\otimes\mathbb{I}\right\} ,\,P_{2}:\left\{ \mathbb{I}\otimes\mathbb{I},X\otimes X\right\} ,\right.$\tabularnewline
 &  &  & $\left.P_{3}:\left\{ \mathbb{I}\otimes\mathbb{I},iY\otimes X\right\} ,\,P_{4}:\left\{ \mathbb{I}\otimes\mathbb{I},iY\otimes\mathbb{I}\right\} \right\} $\tabularnewline
\hline 
4  & 1  & 4-qubit cluster state  & $\left\{ P_{1}:\left\{ \mathbb{I}\otimes\mathbb{I},X\otimes iY\right\} ,\,P_{2}:\left\{ \mathbb{I}\otimes\mathbb{I},X\otimes Z\right\} ,\right.$\tabularnewline
 &  &  & $\left.P_{3}:\left\{ \mathbb{I}\otimes\mathbb{I},iY\otimes Z\right\} ,\,P_{4}:\left\{ \mathbb{I}\otimes\mathbb{I},iY\otimes iY\right\} \right\} $\tabularnewline
\hline 
\end{tabular}
\par\end{centering}

\protect\caption{\label{tab:Multiparty-Quantum-conference.} We present some examples
of the quantum states required for QC and corresponding encoding operations.
In these examples, if one of the party do not encode (consider Identity)
then Protocol 2 will reduce to Protocol 1.}
\end{table}

The proposed QC scheme may also be extended to an asymmetric counterpart
of the QC scheme, where each party may not be encoding the same amount
of information. One such easiest example is a lecture, where the orator
speaks most of the time while the remaining users barely speak. In
such cases, the parties sending redundant bits to accommodate the
QC scheme may choose an AQC scheme. To exploit the maximum benefit
of such schemes a party encoding more information than others (say
Alice) should prepare (and also measure) the quantum state (in other
words, start the QC scheme). In this case, the choice of unitary operations
by each party would also become relevant and Alice should use a subgroup
of higher order than the remaining users. For instance, in a 3-party
scenario, Alice may use a $P_{2}$ from Row 2 of Table \ref{tab:Multiparty-Quantum-conference.}
to encode 2 bits message, while the remaining three users may choose
$P_{1}$ and $P_{3}$, respectively. It is worth noting here that
the security of the QC scheme discussed in the following section ensures
the security of the AQC scheme designed here as well.

Further, the proposed schemes can also be easily modified to obtain
corresponding schemes for controlled QC, where an additional party
(who is referred to as the controller) would prepare the quantum channel
in such a way that the QC task can only be accomplished after the
controller allows the other users to do so \cite{cdsqc,crypt-switch}.
Controlled QC can be achieved in various ways. For example, the controller
may prepare the initial state and keep some of the qubits with himself,
and in absence of the measurement outcome of the corresponding qubits
the other legitimate parties would fail to accomplish the task \cite{cdsqc}.
The same feat can also be achieved by the controller without keeping
a single qubit with himself by using permutation of particles \cite{crypt-switch}.
Thus, it is easy to generalize the proposed schemes for QC to yield
schemes for controlled QC. Such a scheme for controlled QC would have
many applications. For example, a direct application to that scheme
would be quantum telephone where the controller can be a Telephone
company \cite{telephone} that provides the channel to the respective
users after authentication. Thus, the present scheme can be used to
generalize the scheme proposed in \cite{telephone} and thus to obtain
a scheme for multiparty quantum telephone or quantum teleconference.
Additionally, the multiparty communication schemes proposed here can
be reduced to schemes for secure multiparty quantum computation. Interestingly,
a recently proposed secure multiparty computation scheme designed
for quantum sealed-bid auction task \cite{Our-auction} can be viewed
as a reduction of the Protocol 1 proposed here. Therefore, we hope
that the proposed schemes may also be modified to obtain solutions
of various other real life problems.

\section{{\normalsize{}Security analysis and efficiency \label{sec:Security-analysis}}}

A QC protocol is expected to confront the disturbance attack (or denial
of service attack), the intercept-and-resend attack, the entangle-and-measure
attack, man-in-the-middle attack and Trojan-horse\textbf{ }attack
by implementing the BB84 subroutine strategy (for detail see \cite{Kishore-decoy,AQD}),
which allows senders to insert decoy qubits prepared randomly in $X$-basis
or $Z$-basis in analogy with BB84 protocol and to reveal the traces
of eavesdropping by comparing the initial states of the decoy qubits
with the states of the same qubits after measured by the receivers
randomly using $X$-basis or $Z$-basis. In fact, quantum communication
of all the qubits from one party to other, as mentioned in both the
protocols (for example, in Step 1.3), is performed in a secure manner.
To accomplish the secure communication of message qubits using BB84
subroutine, an equal number of decoy qubits (the number of decoy qubits
are required to be equal to the number of message qubits traveling
through the channel) are inserted randomly in the string of travel
qubits. On the authenticated receipt of this enlarged sequence of
travel qubits, the sender discloses the positions of the decoy qubits
and those qubits are then measured by the receiver randomly in $X$-basis
or $Z$-basis. Subsequent comparison of the initial states and the
measurement outcomes reveals the error rate. If the computed error
rate is obtained below a tolerable limit, then the quantum communication
of message qubits is considered to be accomplished in a secure manner
\cite{nielsen,Kishore-decoy}, and the steps thereafter are followed.
Therefore, the above mentioned attacks on the proposed schemes can
be defeated simply by adding decoy qubits and following BB84 subroutine. 

Further, Bob's intimation by Alice that she has sent her qubits and
Bob's acknowledgment of the receipt of qubits, via an authenticated
classical channel, is necessary to avoid the unwanted circumstances
under which Eve pretends as the desired party. There also exist some
technical procedures to circumvent the Trojan-horse attack (\cite{AQD}
and references therein). As a scheme of QC incorporates multiusers
we have discussed below the security in two scenarios where (1) an
outsider (Eve) attacks the protocol, or (2) an insider (one or some
of the legitimate users) attacks the protocol. Further, all the attacks
and counter measures mentioned in this section are applicable on both
the schemes, unless specified.

\subsection*{{\normalsize{}Outsider's attacks}}

In the \textbf{entangle-and-measure attack}, Eve entangles her qubit
$\alpha\left|0\right\rangle +\beta\left|1\right\rangle $ with the
travel qubit in the channel. Eve can extract the information by performing
the $Z$-basis measurement on her ancillae. To counter this attack,
the decoy states $\left|0\right\rangle $, $\left|1\right\rangle $,
$\left|+\right\rangle $ and $\left|-\right\rangle $ are randomly
inserted and when they are examined for security,\textcolor{red}{{}
}then Eve is detected with probability $\left|\beta\right|^{2}$ when
she attacks $\left|0\right\rangle $ and $\left|1\right\rangle $
states, otherwise the states remain separable for $\left|+\right\rangle $
and $\left|-\right\rangle $. Consequently, the total detection probability
of Eve is $\frac{\left|\beta\right|^{2}}{2}$ taking into account
that the probability of generation of each decoy qubit state is $\frac{1}{4}$.

In the\textbf{ intercept-and-resend attack}, Eve prepares some fresh
qubits and swaps one of her qubits with the accessible qubit in the
channel when $\left(i-1\right)$th user sends it to $i$th user. Thereafter,
Eve retrieves her qubit during their communication from $i$th user
to $\left(i+1\right)$th user and obtains the encoding of $i$th user
by performing a measurement on her qubits. This attack will also be
defended by incorporating decoy qubits. However, Eve may modify her
strategy to measure the intercepted qubits randomly in either the
computational or diagonal basis before sending the freshly prepared
qubits corresponding to the measurement outcomes. It is evident that
Eve's measurement of the decoy qubits will produce disturbance if
she measures in the wrong basis. Let $n$ be the total number of travel
qubits such that $\frac{n}{2}$ are decoy and message qubits each.
Eve intercepts $m$ qubits which will entail both decoy and message.
Without a loss of generality, we assume that half of the $m$ qubits
are decoy and the other half are message qubits. Since the security
check is performed on the decoy qubits alone,\textcolor{red}{{} }we
are interested in the $m/2$ decoy qubits which Eve measures in her
lab out of the $\frac{n}{2}$ decoy qubits in the channel. The fraction
of qubits measured by Eve out of the total decoy qubits is given by
$f=\frac{m/2}{n/2}=\frac{m}{n}$. From which the information gained
by Eve is $I(A:E)=f/2.$ This implies that $f/2$ times the correct
basis will be chosen by Eve. The error induced by Eve is observed
by Alice and Bob only when Bob measures in the same basis as of Alice
and is $e=\frac{f/2}{2}=\frac{f}{4}$. The amount of information Bob
receives is given by $I(A:B)=(1-H[\frac{f}{4}])$, where $H\left[u\right]$
is the Shannon binary entropy. The security is ensured until $I(A:B)\geq I(A:E)$.
One can calculate the fraction $f\cong0.68$ for secure communication
with the tolerable error rate $17\%$ (\cite{Err-Rate,Err-Rate2}
and references therein). Eve's success probability is $\frac{3}{4}$
and it would decrease with the increasing value of $m$ as $\left(\frac{3}{4}\right)^{m}$. 

\textbf{Information leakage attack} is inherent in the QD schemes,
and consequently, is applicable to Protocol 2 proposed here as well.
It refers to the information gained by Eve about the encoding of the
legitimate parties by analyzing the classical channel only. In brief,
the leakage can be thought of as the difference between the total
information sent by both the legitimate users and the minimum information
required by Eve to extract that information (i.e., Eve's ignorance).
The mathematical prescription for an average gain of Eve's information
is 
\begin{equation}
\begin{array}{lcl}
I\left(A_{i}:E\right) & = & H_{{\rm a\,priori}}-H_{{\rm a\,posteriori}},\end{array}\label{eq:leakage}
\end{equation}
where $H_{{\rm a\,priori}}$ is the total classical information all
the legitimate parties have encoded; and $H_{{\rm a\,posteriori}}$
is Eve's ignorance after the announcement of the measurement outcome
and is averaged over all the possible measurement outcomes as $\sum_{r}P\left(r\right)H\left(i|r\right)$,
with the conditional entropy $H\left(i|r\right)=-\sum_{i}P\left(i|r\right)\,\log\left(P\left(i|r\right)\right)$.
If the party authorized to prepare and measure the quantum state selects
the initial state randomly and sends it to all the remaining users
by using a standard unconditionally secure protocol for QSDC or DSQC
then the leakage can be avoided as it increases the $H_{{\rm a\,posteriori}}$,
and thus decreases $I\left(A_{i}:E\right)$ to zero corresponding
to no leakage \cite{AQD}.

\subsection*{{\normalsize{}Insider's attacks}}

\textbf{Participant attack }is possible in both the schemes proposed
here.\textbf{ }In the first scheme, a participant can send different
cbits to different members unless we assume semi-honest parties. Although
this scheme is advantageous in certain applications, like sealed bid
auction (where this attack is detected in post-confirmation steps)
\cite{Our-auction} or where each participant wants to encode different
values to respective participant, but in the conference scenario where
it is required that each participant encodes the same message to all
other participants then this attack is prominent, and it is wise to
follow the second scheme, which is free from the assumption of semi-honest
parties. 

In the second scheme, the authorized party (authorized to prepare
and measure the quantum state) encodes his information at the end
just before performing the joint measurement and announcing the outcome.
If he wants to cheat he can disclose an incorrect measurement outcome
corresponding to his modified encoding once he comes to know others'
encoding. This action can be circumvented, and we can implement this
protocol either with a trusted party or we can randomly select any
two participants and run the scheme twice considering that respective
party encodes same information. Another solution would be that the
initiator sends the hash value of his message at the beginning to
all the remaining users, and if the hash value of his encoding revealed
at the last do not match with that of the initially sent hash value,
then he had cheated and will be certainly identified. 

\textbf{Collusion attack }is a kind of illegal collaboration of more
than one party who are not adjacent to each other, to cheat other
members of a group to learn their encoding (precisely of those who
are in between them). The proposed schemes are circular in nature.
In this type of an attack, the attackers generate an entangled state
and circulate the same number of fake qubits as that of the travel
qubits. The attackers at the end already possess the home photons
of the fake qubits circulated by the first attacker and performs a
joint measurement to learn the encoding of the participants in between
them. It will be more effective if $i$th and $\left(i+\frac{n}{2}\right)$th
participants collude. This is so, as both of them get the access of
the travel particles at least once after knowing the secret of all
the remaining parties. This attack can be averted by breaking the
larger circle into $l$ sub-circles such that if less than $l$ attackers
collude, they will not be able to cheat (see \cite{Our-auction} for
details). This attack and the solution are applicable in both the
proposed schemes.

\subsection*{{\normalsize{}Qubit efficiency: }}

The qubit efficiency of a quantum communication scheme is calculated
as 
\[
\eta=\frac{c}{q+b},
\]
where $c$ bits of classical information is transmitted using $q$
number of qubits, and an additional classical communication of $b$
bits \cite{defn of qubit efficiency}. In the first QC scheme, $c=Nk,$
$q=\left(n+mN\right)N$, and $b=0$ as each party sends $k$ bits
and prepares $n$-qubit entangled state and $m$ decoy qubits in each
round of quantum communication. Therefore, the efficiency is calculated
to be $\eta_{{\rm Protocol}\,1}=\frac{k}{\left(mN+n\right)}$.

Similarly, the qubit efficiency of the second QC scheme among $N$
parties such that each party encodes $k$ bits can be computed by
noting that in this case $c=Nk,$ $q=n+Nm$, and $b=n$. Here, $b\neq0,$
as the classical communication of $n$ cbits is associated with the
broadcast of the measurement outcome by the authorized party. Thus,
the qubit efficiency is obtained as $\eta_{{\rm Protocol}\,2}=\frac{k}{m+\left(2n/N\right)}$.
From the $\eta_{{\rm Protocol}\,2}$ one can easily calculate the
qubit efficiency of various possible QC schemes detailed in Table
\ref{tab:conference}. For example, one can check that the qubit efficiency
of a two party QC with each party encoding 2 bits (which is Ba An's
QD protocol) using Bell state as quantum channel is 67\%. Similarly,
the qubit efficiency for a QC scheme involving three parties sending
1 bit each with Bell state as the quantum channel can be obtained
as 43\%. Hence, we find that for the same initial state as quantum
channel the efficiency decreases as the number of parties increases
and/or\textcolor{red}{{} }the number of encoded bits decreases.

\section{{\normalsize{}Conclusion \label{sec:Conclusion}}}

In summary, the notion of QC is introduced as a multiparty secure
quantum communication task which is analogous with the notion of classical
conference, and two protocols for secure QC are designed. The proposed
protocols are novel in the sense that they are the first set of protocols
for QC, as the term QC used earlier were connected to communication
tasks that were not analogous to classical conference. Further, it
is shown that protocols proposed here can be reduced to protocols
for QC proposed earlier considering much weaker notion of conference.
One of the proposed protocols can be viewed as a generalization of
the ping-pong protocol for QSDC, whereas the other one can be viewed
as a generalization of the schemes for QD. It is noted that Protocol
1 composes number of rounds of multiple-sender to single receiver
secure direct communication, which accomplishes the task of QC under
the assumption of semi-honesty of the users. However, this semi-honesty
assumption is not required for Protocol 2, which is proposed here
as multiple-sender to multiple-receiver scheme, where the task is
performed in a single round. Subsequently, both the proposed schemes
are elaborated with the help of an explicit example. 

We have discussed the utility and applications of these protocols
in different scenarios. Specifically, the proposed schemes may be
reduced to a set of multi-party QKD and QKA schemes, if the parties
involved in QC send random bits instead of meaningful messages. Further,
feasibility and significance of the controlled and asymmetric counterparts
of the proposed QC schemes have also been established. The modified
versions of the proposed schemes may also be found useful in accomplishing
some real-life problems, whose primitive is secure multiparty computation.
For example, one can employ the proposed schemes for voting among
the five countries having power of veto in United Nations, where it
is desired that the choice of a voter is not influenced by the choice
of the others. The proposed scheme can also be extended to obtain
a dynamic version of QC, where a participant can join the conference
once it has started and leave it before its termination. Such a generalization
is possible using the method introduced by some of the present authors
in Ref. \cite{HDQSS}. Further, the effect of various types of Markovian
and non-Markovian noise on the schemes proposed here can be investigated
easily using the approach adopted in \cite{Vishal,non-Mar}.

Security of the proposed schemes has been established against various
types of insider and outsider attacks. Further, the qubit efficiency
analysis established that Protocol 2 is more efficient than Protocol
1. Further, one can easily observe that the proposed schemes are much
more efficient compared to a simple minded scheme that performs the
same task by using multiple two-party direct communication schemes,
which will again work only under the assumption of semi-honest users. 

Finally, we have also presented a set of encoding operations suitable
with a host of quantum channels for performing the QC schemes for
number of parties. This provides experimentalists a freedom to choose
the encoding operations and the quantum state to be used as quantum
channel as per convenience. Further, experimental realization of quantum
secure direct communication scheme, which can demonstrate protocols,
like quantum dialogue, quantum authentication, has been successfully
performed in \cite{QSDC}, and it paves way for experimental realization
of QC. Keeping these facts in mind, we conclude this paper with a
hope that the schemes proposed here and/or their variants will be
realized in the near future.

\textbf{Acknowledgment:} AB acknowledges support from the Council
of Scientific and Industrial Research, Government of India (Scientists'
Pool Scheme). CS thanks Japan Society for the Promotion of Science
(JSPS), Grant-in-Aid for JSPS Fellows no. 15F15015. KT and AP thank
Defense Research \& Development Organization (DRDO), India for the
support provided through the project number ERIP/ER/1403163/M/01/1603.

\end{document}